\documentclass[%
reprint,
 amsmath,amssymb,
aps,
]{revtex4-1}

\usepackage{graphicx}
\usepackage{dcolumn}
\usepackage{bm}

\begin{document}

\title{Nonmonotonic bias dependence of local spin accumulation signals\\ in ferromagnet/semiconductor lateral spin-valve devices}

\author{Y. Fujita,$^{1}$$\footnote{Present address: Research Center for Magnetic and Spintronic Materials, National Institute for Materials Science (NIMS), 1-2-1 Sengen, Tsukuba, Ibaraki 305-0047, Japan.}$  M. Yamada,$^{1}$ M. Tsukahara,$^{1}$ T. Naito,$^{1}$ S. Yamada,$^{2,1}$ K. Sawano,$^{3}$ and K. Hamaya$^{2,1}$\footnote{E-mail: hamaya@ee.es.osaka-u.ac.jp}}
\affiliation{
$^{1}$Department of Systems Innovation, Graduate School of Engineering Science, Osaka University, 1-3 Machikaneyama, Toyonaka 560-8531, Japan.}
\affiliation{
$^{2}$Center for Spintronics Research Network, Graduate School of Engineering Science, Osaka University, 1-3 Machikaneyama, Toyonaka 560-8531, Japan.}
\affiliation{
$^{3}$Advanced Research Laboratories, Tokyo City University, 8-15-1 Todoroki, Tokyo 158-0082, Japan.
}

\date{\today}

\begin{abstract}
We find extraordinary behavior of the local two-terminal spin accumulation signals in ferromagnet (FM)/semiconductor (SC) lateral spin-valve devices.
With respect to the bias voltage applied between two FM/SC Schottky tunnel contacts, the local spin-accumulation signal can show nonmonotonic variations, including a sign inversion. 
A part of the nonmonotonic features can be understood qualitatively by considering the rapid reduction in the spin polarization of the FM/SC interfaces with increasing bias voltage. 
In addition to the sign inversion of the FM/SC interface spin polarization, the influence of the spin-drift effect in the SC layer and the nonlinear electrical spin conversion at a biased FM/SC contact are discussed. 

\end{abstract}


\maketitle
\section{Introduction}
The detection of a pure spin current, i.e., the flow of spin angular momentum without a charge current, with spin-precession signals in semiconductors (SCs) has been reported through the measurement of nonlocal voltages \cite{Johnson,Jedema_Nature} in four-terminal lateral spin-valve (LSV) devices with SCs, such as GaAs \cite{Lou_NatPhys,Ciorga_PRB,Uemura_PRB}, InGaAs \cite{Uemura_APEX}, GaN \cite{Bhattacharya_APL}, Si \cite{Jonker_APL,Suzuki_APEX,Ishikawa_PRB,Jansen_PRAP}, Ge \cite{Zhou_PRB,Fujita1,Fujita2,Yamada_APEX}, and SiGe \cite{Naito_APEX}. 
Nonlocal measurements \cite{Johnson,Jedema_Nature} are important to demonstrate reliable spin transport and to investigate spin relaxation phenomena in SCs \cite{Peterson_PRB,Ishikawa_PRB,Fujita1,Fujita2,Hamaya}. 
On the other hand, the transport of spin-polarized charge currents flowing between two ferromagnets (FMs) through SCs also needs to be understood for SC spintronic applications \cite{Dery_Nature, Zutic_RMP,Bratkovsky,Jansen,Ramsteiner,Ciorga}.
To date, there have several reports on the electrical detection of the transport of spin-polarized charge carriers using local two-terminal spin-transport measurements in FM--SC--FM structures \cite{Mattana_GaAs, Appelbaum_Si,Hamaya_PRBR,Sasaki_APL,Ciorga_AAD,Bruski_APL,PLi_PRL,Sasaki_APL2014,Saito_JAP,Kawano_PRmat,Oltscher_2DEG}.
However, because only a few local spin signals have been discussed by a simultaneous comparison with nonlocal spin transport signals in SC-based LSV devices, some of the physics relevant to the magnitude of the local two-terminal spin signals is unclear \cite{Ciorga_AAD,Saito_JAP,Sasaki_APL, Bruski_APL,Oltscher_2DEG,Herfort_PRB,Yamada_SST}. 

According to one-dimensional spin diffusion models \cite{TakahashiMaekawa,Jedema_PRB, Kimura_Metal}, the magnitude of the local spin signal is twice as large as that of the nonlocal spin signal.
For all metallic LSV devices, most of the local spin signals can be explained theoretically by conventional models \cite{Jedema_PRB, Kimura_Metal, KimuraHamaya}.
On the other hand, the correlation between local and nonlocal spin signals is not straightforward in SC-based LSV devices \cite{Sasaki_APL2014,Saito_JAP,Sasaki_APL, Bruski_APL}. Sasaki {\it et al.} \cite{Sasaki_APL} and Bruski {\it et al.} \cite{Bruski_APL} showed that the magnitude of local spin signals is relatively large (4 $\sim$ 10 times) compared to the theoretical values in Si- and GaAs-based LSV devices. 
They consider that this is due to an enhancement of the spin transport length of the SC layers at finite bias voltages \cite{Sasaki_APL, Sasaki_APL2014, Bruski_APL}. Yu {\it et al}. suggested, based on a theoretical study, the presence of a spin-drift effect in the nondegenerate SC layers in FM--SC hybrid systems \cite{Yu}. 
However, because the previous studies on Si \cite{Sasaki_APL,Sasaki_APL2014} used strongly degenerate SC layers and FM/MgO/SC tunnel contacts with non-Ohmic electrical properties, the effect of the bias voltage on the local spin signals remains an open question. At least, the influence of the FM/SC interfaces on the detection of the local spin signals should be discussed in FM--SC hybrid systems. 

Here, we experimentally study the magnitude of the local spin-accumulation signals as a function of the bias voltages applied between the two ferromagnetic contacts in FM--SC LSV devices. 
The LSV devices studied consist of a spin injector and detector with relatively low resistance area products ($RA$) and degenerate Ge as a spin transport layer \cite{Fujita1,Fujita2}, where Ge is an important semiconductor material in the field of spin-related photonics \cite{Jamet_PRB,Pezzoli_SR} and quantum computing \cite{Watzinger_NC} applications. 
We find nonmonotonic variations, including sign inversion, of the local spin-accumulation signals with respect to the bias voltage applied between the two FM/SC contacts. 
A possible mechanism and other important aspects for understanding the local spin-accumulation signals are discussed.  

\section{Experimental}
To explore the local spin signals in FM--SC hybrid systems, we have prepared LSV devices with an $n$-type Ge spin-transport channel and two ferromagnetic contacts, as shown in Fig. 1(a). 
First, an undoped Ge(111) layer ($\sim$28 nm) (LT-Ge) was grown at 350$^\circ$C on a commercial undoped Si(111) substrate ($\rho$ $\sim$ 1000 $\Omega$cm), followed by an undoped Ge(111) layer ($\sim$70 nm) grown at 700$^\circ$C (HT-Ge), where we utilized the two-step growth technique by molecular beam epitaxy (MBE) \cite{Sawano_TSF}.
Next, a 70-nm- or 140-nm-thick phosphorus (P)-doped $n^{+}$-Ge(111) layer (doping concentration $\sim$ 10$^{19}$ cm$^{-3}$) was grown on top by MBE at 350$^\circ$C, as the spin transport layer. The room-temperature carrier concentration of the spin transport layer is 8.2 $\times$ 10$^{18}$ cm$^{-3}$, estimated from Hall effect measurements \cite{Fujita1,Fujita2,Hamaya}. 
To promote tunneling conduction at the FM/Ge interfaces, a P $\delta$-doped Ge layer with an ultra-thin Si layer was grown on top of the $n$$^{+}$-Ge layer \cite{MYamada_APL}.
We have so far developed Schottky-tunnel contacts with a $\delta$-doping layer near the FM/SC interfaces \cite{Ando_APL,Kasahara_JAP}. 
As a spin injector and detector, we grew Co$_{2}$FeAl$_{x}$Si$_{1-x}$ (CFAS) layers \cite{Fujita2}, which is a highly spin-polarized Heusler alloy \cite{Ramsteiner,Hono,Inomata}, on top by nonstoichiometric growth techniques with Knudsen cells by MBE \cite{Hamaya_PRL,SYamada_APL,Fujita2}.
Although atomically smooth heterointerfaces between CFAS and Ge were confirmed, the slight outdiffusion of Ge atoms into the CFAS layer was observed near the CFAS/Ge interface region ($\sim$3 nm) by the high angle annular dark field (HAADF) scanning transmission electron microscopy (STEM) imaging and energy dispersive X-ray spectroscopy (EDS) \cite{Vlado1,Vlado2}. 
Like in our previous works \cite{Fujita1,Fujita2}, the FM/$n$$^{+}$-Ge contacts enabled Schottky tunnel conduction of electrons for electrical spin injection and detection. 

Finally, the grown layers were patterned into contacts with a width of 0.4 $\mu$m (FM1) or 1.0 $\mu$m (FM2). 
The detailed fabrication processes of the LSV devices are presented in Fig. S1 in the Supplemental Material \cite{Supplemental}. 
Device A has a channel width ($w$) of 5.0 $\mu$m and a center-to-center distance ($L$) between the FM contacts of 2.7 $\mu$m. 
A top view of the actual device is shown in Fig. 1(b). 
Device B has $w$ = 7.0 $\mu$m and $L$ = 1.10 $\mu$m (not shown here).
As a reference device, we also fabricated device C, annealed at 300$^\circ$C, with a size the same as that of device B.
For devices A, B, and C, the thickness of the spin transport SC layer is 70 nm.
To observe room-temperature signals, we fabricated device D with the same CFAS contacts, $w =$ 7.0 $\mu$m, and $L \sim$ 1.0 $\mu$m.
For device D, the thickness of the spin transport SC layer is 140 nm. 
As depicted in Fig. 1(a), local and nonlocal voltage measurements were carried out in two- and four-terminal schemes, respectively, in the same device \cite{Johnson,Jedema_Nature,TakahashiMaekawa,Jedema_PRB, Kimura_Metal,KimuraHamaya}. In the two-terminal scheme, spin polarized electrons are injected and extracted beneath the FM/SC contacts, leading to nonequilibrium spin accumulation in the SC layer. 
\begin{figure}
\begin{center}
\includegraphics[width=7.5cm]{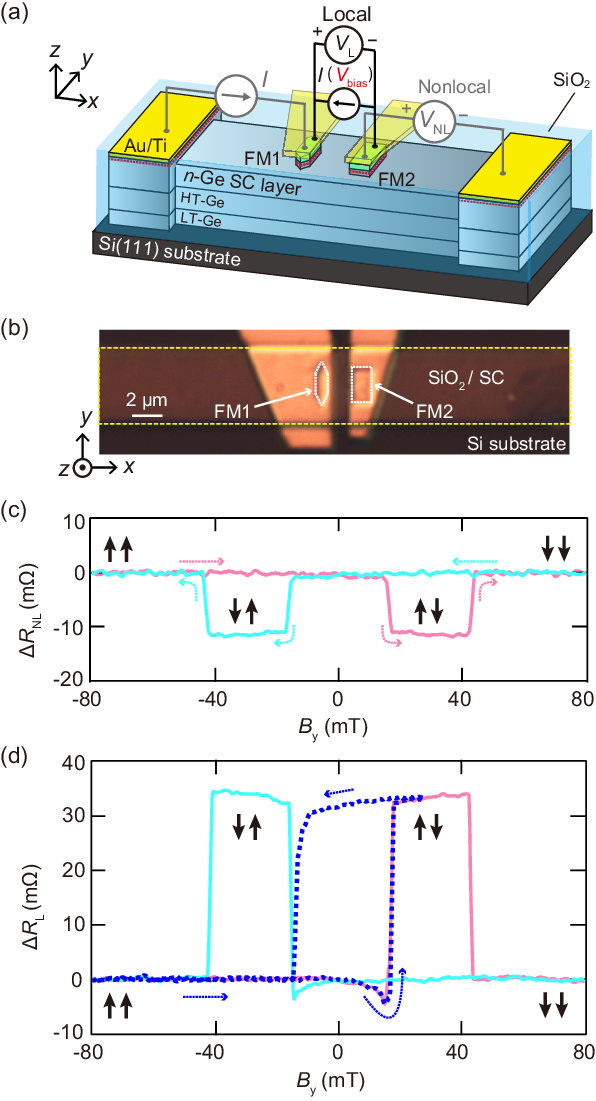}
\caption{(Color online) (a) Schematic illustration of an FM--SC--FM LSV device, showing measurement schemes for nonlocal and local voltage detection. (b) Optical micrograph of an LSV device (device A). (c) Nonlocal magnetoresistance curve measured at $I = -0.5$ mA at 8 K in device A. (d) Local magnetoresistance curve for the same conditions ($I =$ $-$0.5 mA at 8 K). The blue dotted curve is a minor-loop, showing that the anti-parallel magnetization state between FM1 and FM2 is stable. }
\end{center}
\end{figure}
\vspace{5mm}\\

\section{Results and Data Analysis}
\subsection{Spin accumulation signals}
Figure 1(c) shows a representative nonlocal spin signal [$\Delta$$R$$_{\rm NL}$ $=$ $\Delta$$V$$_{\rm NL}$/$I$ $=$ ($V_{\rm NL}^{\uparrow\downarrow} - V_{\rm NL}^{\uparrow\uparrow}$)/$I$] of device A under an in-plane magnetic field ($B_{\rm y}$) at $I =$ $-$0.5 mA at 8 K. 
Here, a negative value of $I$ ($I < 0$) indicates that the spin polarized electrons are injected into the SC from the FM, i.e., a spin injection condition via the Schottky-tunnel barrier.  
For the contacts in device A, $RA$ $\sim$ 200 $\Omega$$\mu$m$^{2}$, which is of the same order as in our previous works \cite{Fujita2}. 
The observed hysteretic nature clearly depends on the parallel and anti-parallel magnetization states between FM1 and FM2, as depicted in the arrows in Fig. 1(c). 
In the nonlocal measurements under an out-of-plane magnetic field ($B_{\rm z}$), we also observed spin-precession signals (Hanle-effect curves), indicating reliable pure spin current transport in the SC layer, as also shown in our previous works \cite{Fujita1,Fujita2,Hamaya}. 
Using the same device (device A), we measured the local spin signal [$\Delta R_{\rm L} = \Delta V_{\rm L}/I$ $=$ ($V_{\rm L}^{\uparrow\downarrow} - V_{\rm L}^{\uparrow\uparrow}$)/$I$] by applying $B_{\rm y}$ under the same conditions ($I =$ $-$0.5 mA at 8 K), as shown in Fig. 1(d). 
Clear positive $\Delta R_{\rm L}$ changes with hysteretic behavior are observed when $B_{\rm y}$ exceeds $\pm$16 mT, meaning that a positive $|{\Delta}R_{\rm L}|$ implies conventional spin-dependent transport of electrons through the SC layer. 
Here, a small negative $\Delta R_{\rm L}$ due to the anisotropic magnetoresistance (AMR) effect in the larger FM electrode (FM2) can be seen within $\pm$16 mT. 
Although this feature cannot be observed in some cases, these AMR signals are proof of the formation of antiparallel states once $B_{\rm y}$ exceeds $\pm$16 mT. 
To verify the reliability, we also plotted minor-loop data, measured under the same conditions, shown as a blue dashed curve. 
The evident minor-loop means that the observed positive $\Delta R_{\rm L}$ changes in Fig. 1(d) can be attributed to the spin-dependent transport of electrons through the SC layer. 
This is proof of the presence of nonequilibrium spin accumulation in the SC layer in FM--SC--FM LSV devices.
In addition, we obtained Hanle-effect curves even in the local measurements by applying $B_{\rm z}$, which is similar to those in the previous works \cite{Sasaki_APL, Spiesser_APL}.
As we focus on the magnitude of the local spin signal $|{\Delta}R_{\rm L}|$ and of the nonlocal spin signal $|{\Delta}R_{\rm NL}|$, the ratio $|{\Delta}R_{\rm L}|$/$|{\Delta}R_{\rm NL}|$ is $\sim$2.7, which is slightly different from the value interpreted in the one-dimensional spin diffusion models \cite{Jedema_PRB, Kimura_Metal}. 
It should be noted that the $|{\Delta}R_{\rm L}|$/$|{\Delta}R_{\rm NL}|$ value is relatively small compared to those in LSV devices with Si \cite{Saito_JAP,Sasaki_APL} and GaAs \cite{Bruski_APL}.

\subsection{Bias voltage effect on spin accumulation}
Figure 2(a) shows $\Delta V_{\rm L}$ versus $B_{\rm y}$ for device A for various $I$ values applied between the two FM contacts at 8 K. 
Interestingly, we can clearly see a sign inversion of $\Delta V_{\rm L}$ even for the same $I$ polarity, indicating that the spin accumulation does not depend linearly on $I$.
To verify this extraordinary behavior, we summarize the detected $\Delta V_{\rm L}$ values as a function of $I$ in Fig. 2(b). 
For both device A and device B, {\it sine-curve like} shapes and sign inversion of $\Delta V_{\rm L}$ for the same $I$ polarity can be seen, resulting in a nonmonotonic variation in $\Delta V_{\rm L}$. 
This behavior has not previously been observed in local two-terminal measurements of FM/SC LSV devices. 
\begin{figure}
\begin{center}
\includegraphics[width=7.5cm]{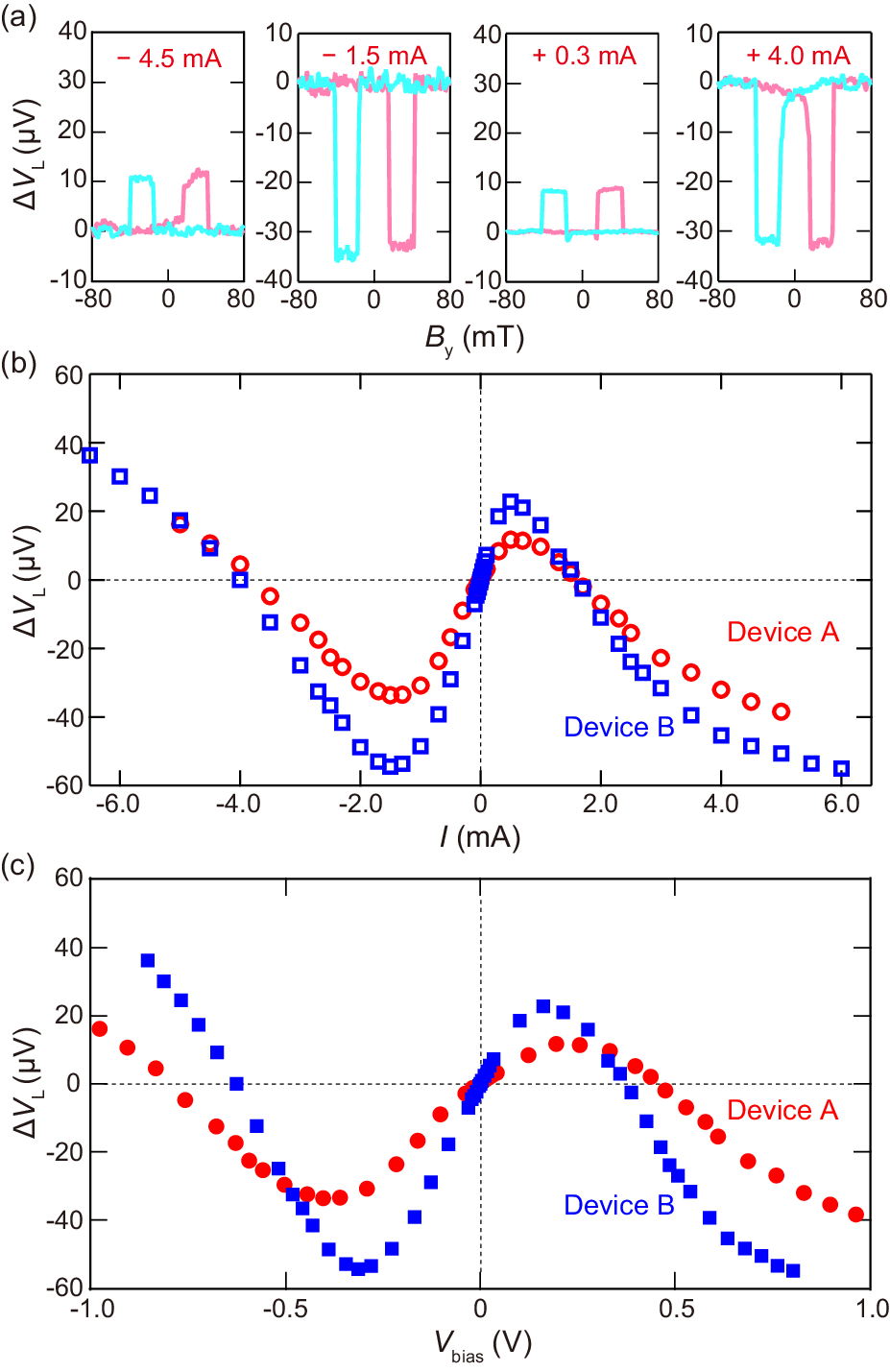}
\caption{(Color online) (a) Local spin accumulation signals at 8 K at $I = -4.5$, $-1.5$, $+0.3,$ and  $+4.0$ mA for device A. 
Sign changes in $\Delta V_{\rm L}$ even for the same $I$ polarity are observed and the magnitude of $\Delta V_{\rm L}$ ($|\Delta V_{\rm L}|$) in the high $I$ region becomes smaller than that in the low $I$ region. (b) $I$ dependence of $\Delta V_{\rm L}$ at 8 K for devices A (open circles) and B (open squares). The amplitude for device B is larger than that for device A because the $L$ value in device B is smaller than that in device A. (c) $V_{\rm bias}$ dependence of $\Delta V_{\rm L}$ at 8 K for devices A (closed circles) and B (closed squares). }
\end{center}
\end{figure}

In the standard theory based on the one-dimensional spin drift-diffusion model in FM1/SC/FM2 systems including double tunnel barriers \cite{Fert_PRB,Fert1,Fert2}, 
$\Delta V_{\rm L}$ increases with increasing magnitude of $I$, and the sign of $\Delta V_{\rm L}$ is associated with the polarity of $I$ as follows: 
\begin{equation}
	\label{eq: Fertmodel}
	\Delta V_{\rm L} =  \frac{8I\gamma_{1}\gamma_{2}{r_{\rm b}^*}^{2}r_{\rm N}}{S\{(2r_{\rm b}^* + r_{\rm N})^{2}\exp(\frac{L}{\lambda_{\rm N}}) - r_{\rm N}^{2}\exp(-\frac{L}{\lambda_{\rm N}})\}},
\end{equation}
where $\gamma_{\rm 1}$ and $\gamma_{\rm 2}$ are the spin polarizations of the FM1/SC and FM2/SC interfaces, $r_{\rm b}^*$ indicates the $RA$ value for the FM/SC interfaces, 
and $\lambda_{\rm N}$, $r_{\rm N}$ and $S$ are the spin diffusion length, the spin resistance, and the cross section area of the SC layer, respectively. 
If $\gamma_{\rm 1}$ and $\gamma_{\rm 2}$ are constant and the spin-dependent transport of electrons through the FM1/SC/FM2 structure stems from the spin accumulation in the SC layer including FM/SC interfaces, the sign of $\Delta V_{\rm L}$ in Eq. (\ref{eq: Fertmodel}) should depend on the polarity of $I$. 
However, the tendency observed in Fig. 2(b) cannot be explained in terms of the change in the polarity of $I$. 
This implies that the data in Fig. 2(b) include the sign inversion of $\gamma_{\rm 1}$ and $\gamma_{\rm 2}$ with increasing magnitude of $I$. 
Sign inversion of the FM/SC interface spin polarization has been presented for some nonlocal LSV systems, such as FM--GaAs--FM \cite{Lou_NatPhys,Salis_PRB,Salis_PRBR2,Hu}. 
In these reports, it has been argued that the sign inversion of the interface spin polarization occurs due to a change of the bias voltage ($V_{\rm bias}$) applied between the two FM contacts \cite{Salis_PRB,Salis_PRBR2,Hu}. 
Thus, to reconsider the behavior in Fig. 2(b) in detail, we summarize $\Delta V_{\rm L}$ as a function of $V_{\rm bias}$, as displayed in Fig. 2(c). 
Fig. 2(c) shows a similar behavior to Fig. 2(b). 
\begin{figure*}
\begin{center}
\includegraphics[width=14cm]{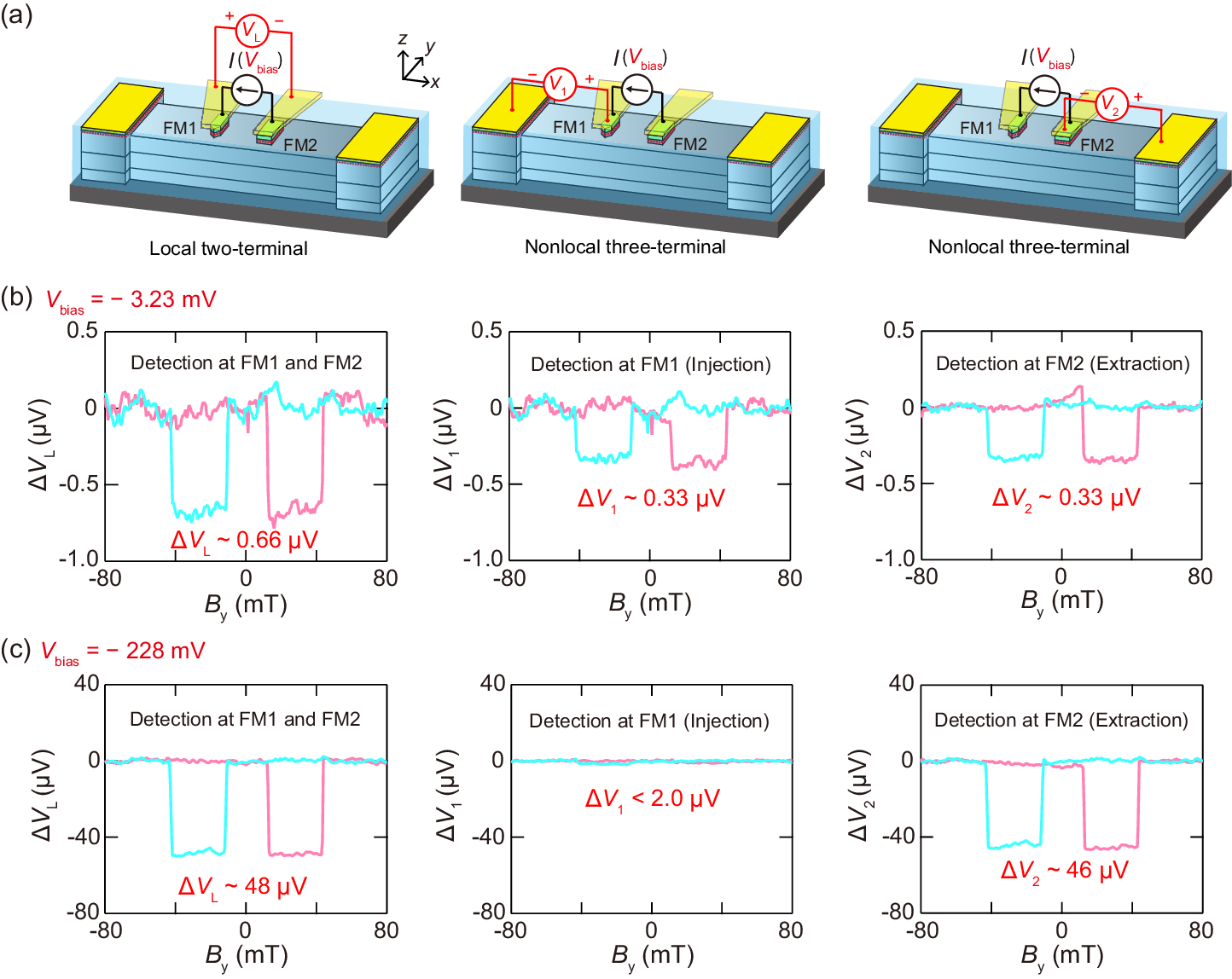}
\caption{(Color online) (a) Schematic illustrations of the geometry for conventional local two-terminal and nonlocal three-terminal measurements. (b) and (c) show the output voltages, $\Delta V_{\rm L}$, $\Delta V_{\rm 1}$, and $\Delta V_{\rm 2}$, versus $B_{\rm y}$ for device B at $V_{\rm bias}$ = $-$3.23 mV and $-$228 mV, respectively. }
\end{center}
\end{figure*}

\subsection{Nonlinear effect and spin-drift effect on spin accumulation}
As described in Sec. III A, the ratio $|{\Delta}R_{\rm L}|$/$|{\Delta}R_{\rm NL}|$ is $\sim$2.7, slightly deviates from the value interpreted in the one-dimensional spin diffusion models \cite{Jedema_PRB, Kimura_Metal}. 
Recently, Jansen {\it el al}. quantitatively clarified that the nonlocal spin accumulation signals in FM--SC--FM LSV devices with tunnel barriers are generally derived from the signals at the spin-detector contacts because a large change in the spin-detection efficiency occurs at biased FM/SC spin-detector contacts \cite{Jansen_PRAP2}. 
Even in local two-terminal or three-terminal measurements, the nonlinear electrical spin conversion at a biased FM/SC spin-detector contact should be considered, as discussed in previous works \cite{Chantis2,Ando_PRB,Crooker_PRB,Shiogai_PRB,Jansen_PRAP2}. 

To investigate the influence of the nonlinear spin detection efficiency at a biased FM/SC spin-detector contact, we made nonlocal three-terminal measurements \cite{Sasaki_APL2014,Jansen_PRAP2}. 
As schematically illustrated in Fig. 3(a), the output voltage change, $\Delta V_{\rm 1}$ or $\Delta V_{\rm 2}$, of the nonlocal three-terminal measurements indicates spin accumulation underneath the FM1 or FM2 contact, respectively, under the application of $V_{\rm bias}$ between the FM1 and FM2 contacts. 
If we use a negative $V_{\rm bias}$, $V_{\rm bias} <$ 0, the electron spins can be injected from the FM1 contact and extracted from FM2. 
When we apply a very low $V_{\rm bias}$ of $-$ 3.23 mV ($I =$ $-$ 0.01 mA), a local spin accumulation voltage of $\Delta V_{\rm L} \sim$ 0.66 $\mu$V can be obtained, as shown in the left figure of Fig. 3(b). 
Under this condition, nonlocal three-terminal measurements reveal that both $\Delta V_{\rm 1}$ and $\Delta V_{\rm 2}$ are $\sim$ 0.33 $\mu$V, which is half the magnitude of $\Delta V_{\rm L}$. 
This feature is different from previous reports that include a large nonlinear effect due to the MgO tunnel barrier \cite{Sasaki_APL2014,Jansen_PRAP2} and FM/GaAs Schottky tunnel barriers \cite{Crooker_PRB,Shiogai_PRB}. 
We can verify that the local spin accumulation signal at $V_{\rm bias}$ = $-$3.23 mV is produced by both the FM1 and FM2 contacts, which can be interpreted within a framework of the standard theory \cite{Jedema_PRB, Kimura_Metal}. In short, even a linear response can appear for a very low $V_{\rm bias}$. 
With increasing $V_{\rm bias}$, on the other hand, the correlation between $\Delta V_{\rm L}$ and $\Delta V_{\rm 1}$ (or $\Delta V_{\rm 2}$) is markedly varied. 
When $V_{\rm bias}$ = $-$228 mV ($I =$ $-$ 1.0 mA) was applied, the total spin accumulation signal detected by the local two-terminal measurement derives mostly from the spin accumulation at the FM2 contact, as shown in Fig. 3(c). This feature is similar to those in previous works \cite{Sasaki_APL2014,Jansen_PRAP2}.
Therefore, the linear and nonlinear effects on the local spin accumulation signals coexist in our FM--SC--FM LSV devices.

In addition, the influence of the spin-drift effect on the magnitude of $\Delta V_{\rm L}$ should be considered \cite{Yu,Sasaki_APL2014}. 
For device B at 8 K, for example, a critical electric field of the spin-drift effect, $E_{\rm crit}$ $=$ $\epsilon_{\rm drift}$/$e\lambda_{\rm N}$, where $\epsilon_{\rm drift}$ is an energy scale given by $eD$/$\mu_{\rm e}$ \cite{Yu,Jansen_PRAP2}, can be roughly estimated to be approximately 110 kV/m, larger than $V_{\rm bias}$ = $\pm$ 0.55 V. 
Thus, we speculate that the spin-drift effect induced by the electric-field applied to the SC channel layers, discussed in Ref. \cite{Sasaki_APL, Bruski_APL}, can be ignored for $|V_{\rm bias}| <$ 0.55 V. 
The magnitude of $\Delta V_{\rm L}$ did not depend linearly on $V_{\rm bias}$ even in the $|V_{\rm bias}| <$ 0.55 V region of Fig. 2(c). Hence we should consider other origins to understand the nonmonotonic variation in $\Delta V_{\rm L}$.  

\subsection{Estimation of the interface spin polarization}
Even though we take into account the nonlinear electrical spin conversion effect at a biased FM contact and the spin-drift effect \cite{Jansen_PRAP2}, the sign inversion of the spin accumulation signals shown in Fig. 2 could not be explained.
As described in Sec. III B, the sign inversion of the FM/SC interface spin polarization by a change in $V_{\rm bias}$ should be considered \cite{Salis_PRB,Salis_PRBR2,Hu}. 
In general, the value of $V_{\rm bias}$ shown in Fig. 2(c) is related to the interface voltages, $V_{\rm int1}$ and $V_{\rm int2}$, applied to the FM1/SC and FM2/SC interfaces, respectively, in addition to the voltage ($V_{\rm SC}$) applied to the SC channel layer in FM1--SC--FM2 LSV devices. 
First, we roughly regard $V_{\rm bias}$ as ($V_{\rm int1} - V_{\rm int2}$) because $V_{\rm SC}$ is relatively small for $|V_{\rm bias}| <$ 0.55 V.
For 0 $< V_{\rm bias} <$ 0.55 V,  we can take the value of $V_{\rm int1}$ ($>$ 0) in a spin extraction condition of the FM1/SC contact and that of $V_{\rm int2}$ ($<$ 0) in a spin injection condition of the FM2/SC contact. 

To evaluate the spin polarizations of the FM1/SC and FM2/SC interfaces, we focus again on the nonlocal four-terminal spin accumulation voltages ($\Delta V_{\rm NL}$) in the same devices. 
Figure 4(a) shows plots of $\Delta V_{\rm NL}$ versus $V_{\rm int1}$ and $V_{\rm int2}$ for devices A and B, where two kinds of $\Delta V_{\rm NL}$ can be obtained by exchanging between the spin injector and detector for each device, and $V_{\rm int1}$ or $V_{\rm int2}$ stands for the bias voltage applied to the FM1/SC or FM2/SC interfaces, respectively, detected by the three-terminal current-voltage measurements, as shown in the inset figures.
For $V_{\rm int1}, V_{\rm int2} < 0$, i.e., spin-injection conditions of electrons from FM to SC, the positive $\Delta V_{\rm NL}$ values increase with increasing $|V_{\rm int1}|$ or $|V_{\rm int2}|$, although those are slightly suppressed only in the high $|V_{\rm int}|$ regime. 
On the other hand, for $V_{\rm int1}$, $V_{\rm int2} > 0$ (spin extraction condition), the enhancement of the negative $\Delta V_{\rm NL}$ values is markedly suppressed, and $\Delta V_{\rm NL}$ approaches zero at around $V_{\rm int1}$, $V_{\rm int2} =$ + 0.3 V. 
These asymmetric features with respect to $V_{\rm int1}$, $V_{\rm int2} =$ 0 lead to the strong nonmonotonicity. 
A similar nonmonotonicity in the nonlocal spin accumulation signals has already been observed in FM--GaAs--FM LSV devices \cite{Salis_PRB,Salis_PRBR2}, and the origin of the asymmetry in $\Delta V_{\rm NL}$ versus the bias voltage applied to the FM/SC interface in Ref.\cite{Salis_PRB,Salis_PRBR2} was discussed based on the change in the  injection/detection efficiencies at the FM/SC contacts. 

\begin{figure*}
\begin{center}
\includegraphics[width=14cm]{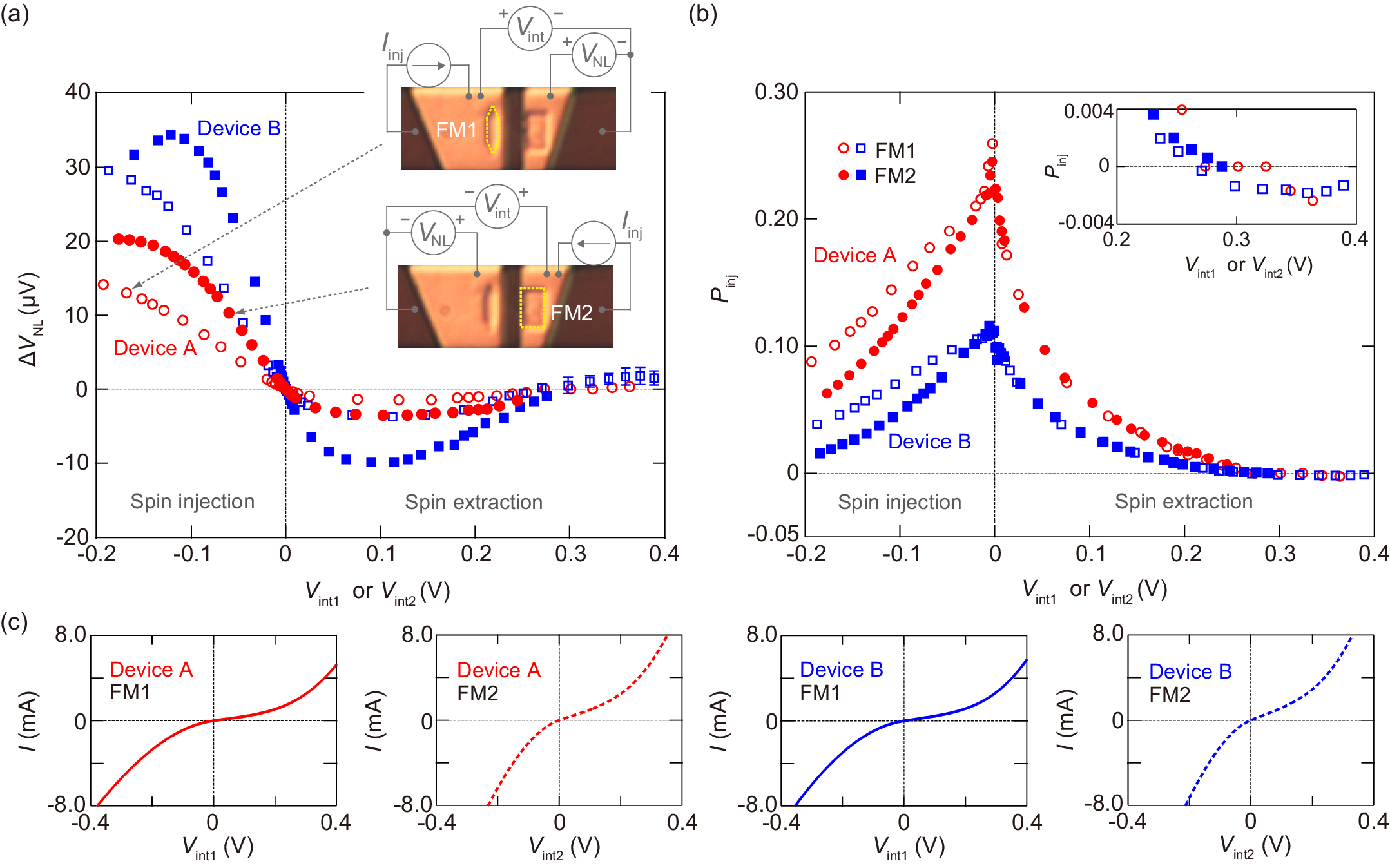}
\caption{(Color online) (a) $V_{\rm int}$ dependence of $\Delta V_{\rm NL}$ at 8 K for devices A (circles) and B (squares). 
The open and closed symbols denote the data for FM1 and FM2, respectively, as a spin injector, and the insets show the measurement schemes. 
(b) $V_{\rm int}$ dependence of the spin polarization ($P_{\rm inj}$) created by FM1 or FM2 in devices A and B. The inset shows an enlarged figure at $V_{\rm int} \sim$ 0.3 V.
(c) $I$--$V_{\rm int}$ characteristics of each FM contact for devices A and B at 8 K. }
\end{center}
\end{figure*}

If we regard the spin polarizations created from the spin injector and spin detector contacts as $P_{\rm inj}$ and $P_{\rm det}$, the correlation among $\Delta V_{\rm NL}$, $P_{\rm inj}$, and $P_{\rm det}$ can be expressed as follows \cite{Lou_NatPhys, Johnson, Jedema_Nature}: 
\begin{equation}
\label{eq: NL}
\Delta V_{\rm NL} ={\frac{{P_{\rm inj}}{P_{\rm det}}{I_{\rm inj}}{\rho_{\rm N}}{\lambda_{\rm N}}}{S}}{\exp\left(-\frac{L}{\lambda_{\rm N}}\right)},
\end{equation}
where $\rho_{\rm N}$ is the resistivity (17.4 $\Omega$$\mu$m) of the SC layer.  The values of $S$ for devices A and B are 0.35 $\mu$m$^{2}$ and 0.49 $\mu$m$^{2}$, respectively. 
For our SC spin transport layer, the value of $\lambda_{\rm N}$ has already been clarified to be 0.56 $\mu$m at 8 K \cite{Yamada_PRB}. 
If the FM1/SC contact is used as a spin injector in the nonlocal voltage measurements, $P_{\rm inj}$ can change with increasing $V_{\rm int1}$. 
On the other hand, the spin polarization of the non-biased contact (FM2/SC), $P_{\rm det}$, can be regarded as being constant. 
Only under very low $V_{\rm int1}$ or $V_{\rm int2}$ conditions can we roughly consider that the assumption $|P_{\rm inj}| = |P_{\rm det}|$ is valid, leading to the values $P_{\rm det} =$ 0.25 and 0.11 for devices A and B, respectively. 
Employing these $P_{\rm det}$ values and the above parameters, we can determine the value of $P_{\rm inj}$ for various $I_{\rm inj}$, which can be converted to $V_{\rm int1}$ or $V_{\rm int2}$. 
The plots of the determined $P_{\rm inj}$ versus $V_{\rm int1}$ or $V_{\rm int2}$ for the FM1/SC and FM2/SC contacts, respectively, in devices A and B are presented in Fig. 4(b). 
With increasing $|V_{\rm int1}|$ or $|V_{\rm int2}|$, $P_{\rm inj}$ decreases, similar to the case of magnetic tunnel junctions \cite{Jansen_PRL,Yuasa_NM}.  
The decrease in $P_{\rm inj}$ for $V_{\rm int1}$, $V_{\rm int2}$ $>$ 0 (spin extraction condition) is slightly larger than that for $V_{\rm int1}$, $V_{\rm int2}$ $<$ 0 (spin injection condition), leading to the asymmetrical bias dependence of $P_{\rm inj}$. 
Because the current--voltage characteristics of the FM/SC Schottky-tunnel contacts used have a small asymmetry with respect to $V_{\rm int1}$, $V_{\rm int2} =$ 0, as shown in Fig. 4(c), the asymmetrical bias dependence of $P_{\rm inj}$ is regarded as a consequence of the asymmetric structure of the energy barrier in the FM/SC Schottky-tunnel contacts. 
For finite $V_{\rm int1}$ and $V_{\rm int2}$, the electronic band structure of FM materials \cite{Jansen_PRL,Valenzuela_PRL,Bruski3} or interfacial states \cite{Hu,Salis_PRB,Salis_PRBR2} can also affect the spin polarization of electrons through the FM/SC interface.
As shown in previous works \cite{Salis_PRB,Salis_PRBR2,Valenzuela_PRL}, the sign inversion of $P_{\rm inj}$ created by the FM1/SC and FM2/SC contacts can be observed at $V_{\rm int1}$, $V_{\rm int2} \sim$ $+$ 0.3 V. 

\subsection{Qualitative reproduction of nonmonotonic behavior}
Tentatively regarding the above $P_{\rm inj}$ values separately estimated for FM1/SC and FM2/SC contacts as $\gamma_{\rm 1}$ and $\gamma_{\rm 2}$ of the FM1/SC and FM2/SC interfaces in Eq. (\ref{eq: Fertmodel}), we can also discuss the local spin accumulation voltage $\Delta V_{\rm L}$ in FM1--SC--FM2 LSV devices. 
Here, $r_{\rm N} (=\rho_{\rm N} \times \lambda_{\rm N}) = $ 9.74 $\Omega$$\mu$m$^{2}$ in our LSV devices and the $r_{\rm b}^*$ values vary within the range of 70 $\Omega$$\mu$m$^{2}$ $\le r_{\rm b}^*$ $\le$ 470 $\Omega$$\mu$m$^{2}$ in the local two-terminal measurement conditions.
From Eq. (\ref{eq: Fertmodel}) under this condition, we can roughly consider the following relation, $\Delta V_{\rm L} \propto$ $\gamma_{1}$$\gamma_{\rm 2}$$I$. 
\begin{figure}[t]
\begin{center}
\includegraphics[width=8.5cm]{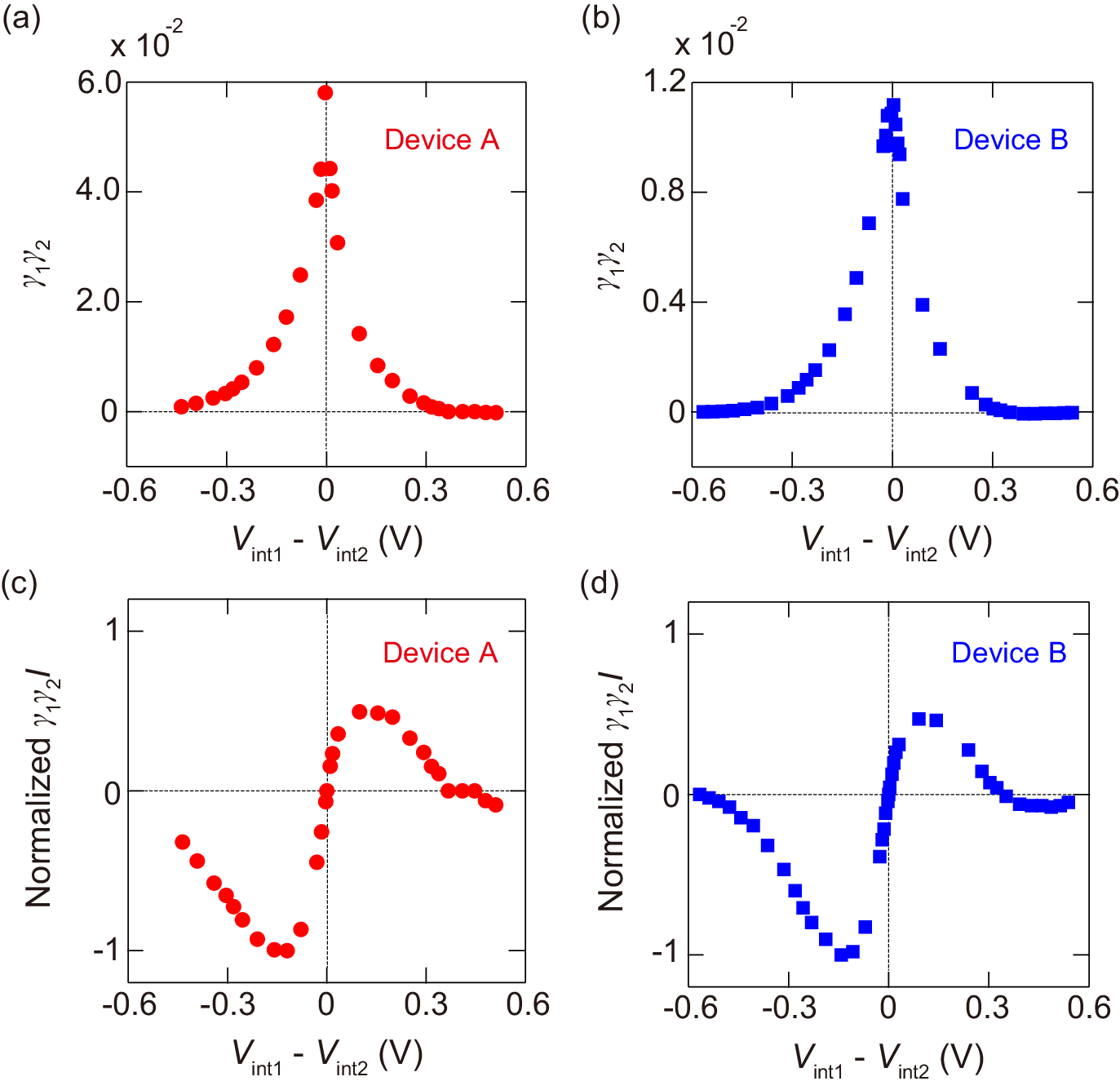}
\caption{(Color online) $\gamma_{\rm 1}$$\gamma_{\rm 2}$ as a function of ($V_{\rm int1}$ $-$ $V_{\rm int2}$) for (a) device A and (b) device B. Normalized $\gamma_{\rm 1}$$\gamma_{\rm 2}$$I$ as a function of ($V_{\rm int1}$ $-$ $V_{\rm int2}$) for (c) device A and (d) device B. }
\end{center}
\end{figure}

When we assume that $V_{\rm bias}$ $>$ 0 ($<$ 0) consists of $V_{\rm int1}$ for spin extraction (spin injection) and $V_{\rm int2}$ for spin injection (spin extraction), we can plot $\gamma_{1}$$\gamma_{\rm 2}$ versus ($V_{\rm int1}$ $-$ $V_{\rm int2}$), as shown in Figs. 5(a) and 5(b), for device A and device B, respectively.
For both devices, $\gamma_{1}$$\gamma_{\rm 2}$ rapidly decreases for $|V_{\rm int1}$ $-$ $V_{\rm int2}|$ $<$ 0.15 V. For $|V_{\rm int1}$ $-$ $V_{\rm int2}|$ $>$ 0.15 V, on the other hand, the decrease in $\gamma_{1}$$\gamma_{\rm 2}$ is slow. 
Using the $\gamma_{1}$$\gamma_{\rm 2}$ data in Figs. 5(a) and 5(b), we can estimate $\gamma_{1}$$\gamma_{\rm 2}$$I$, where $I$ is determined from the data in Fig. 4(c). 
For example, when $V_{\rm int1}$ (extraction) for the FM1/SC contact and $V_{\rm int2}$ (injection) for the FM2/SC contact are $+$0.11 V and $-$0.033 V, respectively, the value of $I$ flowing in the FM1-SC-FM2 structure is estimated to be $+$0.5 mA from the data in Fig. 4(c). 
As a result, we can assign $\gamma_{\rm 1}$ and $\gamma_{\rm 2}$ to 0.024 and 0.095, respectively, resulting in $\gamma_{1}$$\gamma_{\rm 2}$$I \sim$ 0.00114. 
In Figs. 5(c) and 5(d), the normalized values of $\gamma_{1}$$\gamma_{\rm 2}$$I$ versus ($V_{\rm int1}$ $-$ $V_{\rm int2}$) are plotted for device A and device B, respectively. 
These figures clearly show the nonmonotonic variations with respect to the bias voltage.
This means that the feature in Fig. 2(c) is related to the intrinsic bias-dependent $\gamma_{1}$$\gamma_{\rm 2}$, as shown in Figs. 5(a) and 5(b), in our LSV devices. 
Since ($V_{\rm int1}$ $-$ $V_{\rm int2}$) in Fig. 5 deviates from $V_{\rm bias}$ in Fig. 2(c), the influence of $V_{\rm SC}$ on $V_{\rm bias}$ should also be addressed. 
Considering the size and resistivity of the SC channel layers in device A and device B \cite{Spiesser}, we determined that $V_{\rm SC}$ is roughly from 0 to $\pm$ 0.5 V and from 0 to $\pm$0.2 V, respectively. As a result, the nonmonotonic behavior in $\Delta V_{\rm L}$ in Fig. 2(c) could be qualitatively reproduced for $|V_{\rm bias}| <$ 0.55 V. 

\section{Discussion}
Because our LSV devices have relatively low $RA$ values of $\sim$200 $\Omega$$\mu$m$^{2}$ \cite{Fujita2}, we can observe local spin accumulation signals over a relatively wide $V_{\rm bias}$ range compared to previous works \cite{Saito_JAP,Sasaki_APL,Bruski_APL}. Due to this advantage, the nonmonotonic behavior in $\Delta V_{\rm L}$ versus $V_{\rm bias}$ was found, as presented in Fig. 2(c). 
As described in the previous sections, the nonmonotonic variations in $\Delta V_{\rm L}$ can be interpreted qualitatively in terms of the intrinsic feature of the bias-dependent $\gamma_{1}$$\gamma_{\rm 2}$ in the FM1--SC--FM2 LSV devices. 
For comparison, we also show $\Delta V_{\rm L}$ as a function of $V_{\rm bias}$ for device C in Fig. 6(a), where device C was an LSV device with CFAS contacts annealed at 300$^{\circ}$C.
Note that the features of the plot of $\Delta V_{\rm L}$ versus $V_{\rm bias}$ are markedly changed and the sign inversion of $\Delta V_{\rm L}$ disappears, which is very different from Fig. 2(c). 

To discuss the above variations after annealing, $\Delta V_{\rm NL}$ as a function of $V_{\rm int1}$ and $V_{\rm int2}$ for device C was also measured, shown in Fig. 6(b). 
Compared to Fig. 4(a), the asymmetry with respect to $V_{\rm int1}$, $V_{\rm int2} =$ 0 is relatively small for each FM/SC contact, and the magnitude of $\Delta V_{\rm NL}$ becomes small. 
In addition, the sign inversion of $\Delta V_{\rm NL}$ for $V_{\rm int} >$ 0 also disappears in device C. 
Here, although there was no influence of the post annealing at 300$^{\circ}$C on the extracted parameters, such as $\lambda_{\rm N}$, and the spin lifetime of the SC layer used, the degradation of the  FM/SC interface quality was directly clarified by HAADF-STEM imaging \cite{Vlado2}.
This implies that the sign inversion of the FM/SC interface spin polarization in our LSV devices is associated with the quality of the FM/SC interface at least.
In many other hybrid systems such as FM/GaAs \cite{Lou_NatPhys,Crooker,Schultz,Hu} and FM/h-BN/graphene \cite{Dash_SR,Gurram_NC}, the sign inversion of the interface spin polarization enables modulation of the spin-related output signals. Even for such cases, some explanations, such as the presence of the resonant states \cite{Hamaya_PRBR,Hu,Tsymbal_PRL} and the density of states of the FM material \cite{Valenzuela_PRL,Bruski3}, have been discussed to explain the sign inversion of the spin polarization at FM/SC interfaces. 
We infer that the spin polarization at the FM/SC interfaces in our LSV devices can be inverted by applying a bias voltage. 

Employing the same data analysis described in Sec. III E, we can roughly obtain the plot of the normalized $\gamma_{1}$$\gamma_{\rm 2}$$I$ versus ($V_{\rm int1}$ $-$ $V_{\rm int2}$) for device C, as shown in the inset of Fig. 6(b). The obtained feature is similar to that in Fig. 6(a), implying that the data analysis is qualitatively useful. 
For these reasons, the nonmonotonic behavior of $\Delta V_{\rm L}$ in Fig. 2(c) can be understood qualitatively by considering the rapid reduction in the spin polarization of the FM/SC interfaces with increasing bias voltage, shown in Figs. 5(a) and 5(b).  

\begin{figure}[t]
\begin{center}
\includegraphics[width=7.5cm]{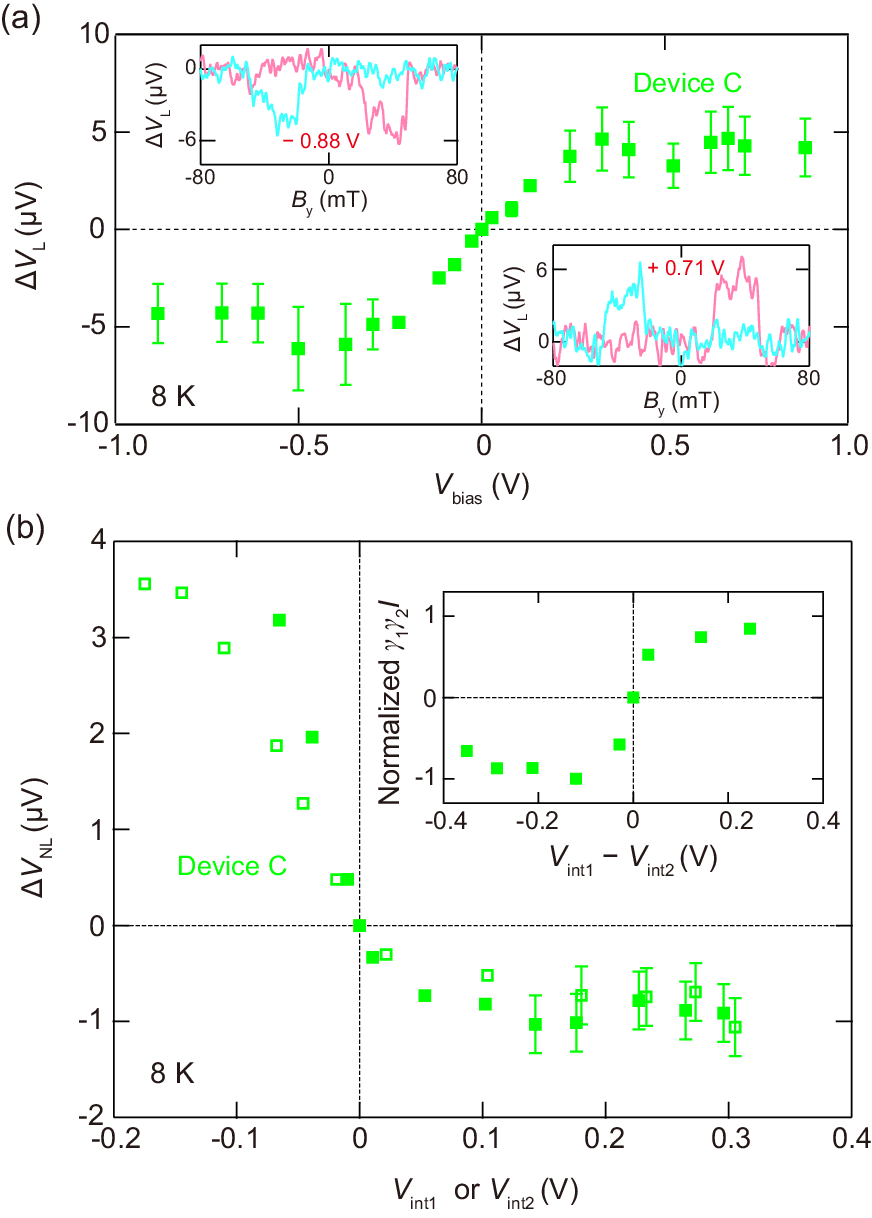}
\caption{ (Color online) (a) $V_{\rm bias}$ dependence of $\Delta V_{\rm L}$ for device C at 8 K. The inset figures show representative local spin accumulation signals at 8 K at $V_{\rm bias} =$ $-$0.88 and +0.71 V.  (b) $V_{\rm int}$ dependence of $\Delta V_{\rm NL}$ at 8 K for device C. The open and closed symbols denote the data for FM1 and FM2, respectively. A plot of normalized $\gamma_{\rm 1}$$\gamma_{\rm 2}$$I$ versus ($V_{\rm int1}$ $-$ $V_{\rm int2}$) is shown in the inset.}
\end{center}
\end{figure}

Because we could not obtain the wide-range temperature dependence of the local spin accumulation signals for devices A and B, we used device D with a smaller $L$ ($L \sim$ 1.0 $\mu$m, $RA \sim$ 100 $\Omega$$\mu$m$^{2}$). 
As shown in the inset of Fig. 7(a), we can observe the local magnetoresistance in an SC-based LSV device with FM/SC Schottky tunnel contacts at room temperature \cite{Tsukahara_APEX}. 
Figure 7(a) shows $\Delta V_{\rm L}$ as a function of $V_{\rm bias}$ up to room temperature (296 K) for device D. 
Note that a similar nonmonotonic behavior shown in Fig. 2(c) can be observed from 150 to 296 K, indicating the reproducibility of the nonmonotonic bias dependence of spin accumulation up to room temperature. However, the sign inversion phenomenon in $\Delta V_{\rm L}$ gradually disappears with increasing temperature. 
In our previous work \cite{Tsukahara_APEX}, it was verified that the interface spin polarization of FM/SC contacts decreases with increasing temperature.
Therefore, we can conclude that the appearance of the sign inversion of $\Delta V_{\rm L}$ is also related to the interface spin polarization of the FM/SC contacts. 
\begin{figure}[t]
\begin{center}
\includegraphics[width=8cm]{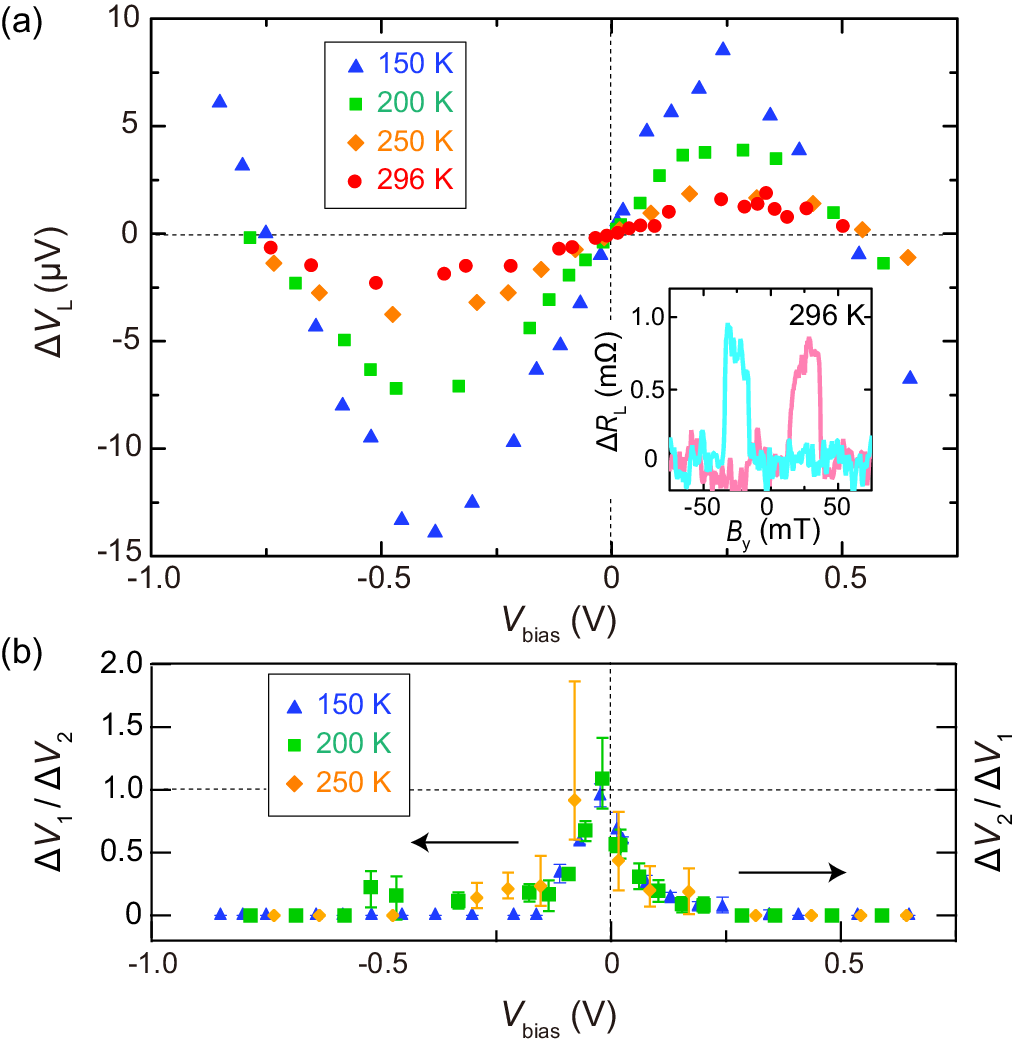}
\caption{(Color online) (a) $V_{\rm bias}$ dependence of $\Delta V_{\rm L}$ from 150 to 296 K for device D. The inset shows a representative local magnetoresistance curve observed at 296 K. (b) $V_{\rm bias}$ dependence of $\Delta V_{\rm 1}$/$\Delta V_{\rm 2}$ ($V_{\rm bias} <$ 0) and $\Delta V_{\rm 2}$/$\Delta V_{\rm 1}$ ($V_{\rm bias} >$ 0), measured by the nonlocal three-terminal method at 150, 200, and 250 K. }
\end{center}
\end{figure}

However, the origin of the salient sign inversion of $\Delta V_{\rm L}$, such as for $|V_{\rm bias}| >$ 0.55 V in Fig. 2(c), could not be precisely identified. 
We finally reconsider the influence of the spin-drift effect in the SC channel layer \cite{Yu,Spiesser} and the nonlinear electrical spin conversion at a biased FM/SC contact \cite{Jansen_PRAP2}. 
As described in Sec. III C, we should take into account the presence of the spin-drift effect in $|V_{\rm bias}| >$ 0.55 V. 
The negative interface spin polarization can be enhanced by the spin-drift effect for $|V_{\rm bias}| >$ 0.55 V in local two-terminal measurements. Further quantitative investigations should be conducted  \cite{Spiesser} . 
On the other hand, for the data in Fig. 7(a), we discuss the influence of the nonlinear effect at a biased FM/SC contact on the local spin accumulation signals in a wide temperature range. 
Figure 7(b) displays plots of $\Delta V_{\rm 1}$/$\Delta V_{\rm 2}$ ($V_{\rm bias} <$ 0) and $\Delta V_{\rm 2}$/$\Delta V_{\rm 1}$ ($V_{\rm bias} >$ 0) versus $V_{\rm bias}$ at various temperatures, where $\Delta V_{\rm 1}$ and $\Delta V_{\rm 2}$ were recorded in nonlocal three-terminal measurements, as shown in Fig. 3. 
Unfortunately, because of the large electrical noise, we could not show reliable data at 296 K. 
If the magnitude of $\Delta V_{\rm 1}$/$\Delta V_{\rm 2}$ ($V_{\rm bias} <$ 0) or $\Delta V_{\rm 2}$/$\Delta V_{\rm 1}$ ($V_{\rm bias} >$ 0) is equal to 1.0, the local spin accumulation signal, $\Delta V_{\rm L}$, can be explained only in terms of the standard theory \cite{Jedema_PRB, Kimura_Metal}. 
In short, the deviation from 1.0 indicates a practical influence of the nonlinear electrical spin conversion at a FM1/SC or FM2/SC contact \cite{Jansen_PRAP2}, as discussed in Sec. III C. 
As can be seen in Fig. 7(b), the nonlinear effect at the FM1/SC or FM2/SC contact on $\Delta V_{\rm L}$ becomes significant, apart from around $V_{\rm bias} \sim$ $-$2 mV. 
Therefore, in addition to the sign inversion of the interface spin polarization, the spin-drift effect and nonlinear electrical spin conversion at the FM1/SC or FM2/SC contact cannot be ignored when explaining the large deviation between the experimental and calculated data for $|V_{\rm bias}| >$ 0.55 V in Fig. 2(c). 
Further theoretical discussion is required to completely understand the nonmonotonic behavior with sign inversion in Fig. 2(c). 

\section{Conclusion}
We found extraordinary behavior of local spin-accumulation signals in FM--SC--FM LSV devices. 
With respect to the bias voltage applied between the two FM/SC contacts, the local spin-accumulation signal showed nonmonotonic variations including sign inversion. 
A part of the nonmonotonic features can be understood qualitatively by considering the rapid reduction in the spin polarization of the FM/SC interfaces. 
In addition to the sign inversion of the FM/SC interface spin polarization, the influence of the spin-drift effect in the SC layer and the nonlinear electrical spin conversion at a biased FM/SC contact should be considered. 

\section{Acknowledgements}
Y.F. and K.H. acknowledge  Dr. Ron Jansen of AIST in Japan for useful discussion to interpret the experimental data.  
This work was partly supported by Grants-in-Aid for Scientific Research (A) (No. 16H02333) and (S) (No. 17H06120 and 19H05616) from the Japan Society for the Promotion of Science (JSPS). 
Y.F. and M.Y. acknowledge JSPS Research Fellowships for Young Scientists (No. 14J03484 and 18J00502).


\begin{thebibliography}{99} 
\bibitem{Johnson}
M. Johnson and R. H. Silsbee, Phys. Rev. Lett. {\bf 55}, 1790 (1985).
\bibitem{Jedema_Nature}%
F. J. Jedema, H. B. Heersche, A. T. Filip, J. J. A. Baselmans, and B. J. van Wees, Nature (London) {\bf 416}, 713 (2002).
\bibitem{Lou_NatPhys}
X. Lou, C. Adelmann, S. A. Crooker, E. S. Garlid, J. Zhang, K. S. M. Reddy, S. D. Flexner, C. J. Palmstr{\o}m, and P. A. Crowell, Nat. Phys. {\bf 3}, 197 (2007).
\bibitem{Ciorga_PRB}%
M. Ciorga, A. Einwanger, U. Wurstbauer, D. Schuh, W. Wegscheider, and D. Weiss, Phys. Rev. B {\bf 79}, 165321 (2009).
\bibitem{Uemura_PRB}
T. Akiho, J. Shan, H. X. Liu, K. I. Matsuda, M. Yamamoto, and T. Uemura, Phys. Rev. B {\bf 87}, 235205 (2013).
\bibitem{Uemura_APEX}
T. Akiho, M. Yamamoto, and T. Uemura, Appl. Phys. Express {\bf 8}, 093001 (2015).
\bibitem{Bhattacharya_APL} 
A. Bhattacharya, M. Z. Baten, and P. Bhattacharya, Appl. Phys. Lett. {\bf 108}, 042406 (2016). 

\bibitem{Jonker_APL}
O. M. J. van't Erve, A. T. Hanbicki, M. Holub, C. H. Li, C. Awo-Affouda, P. E. Thompson, and B. T. Jonker,  Appl. Phys. Lett. {\bf 91}, 212109 (2007).
\bibitem{Suzuki_APEX}
T. Suzuki, T. Sasaki, T. Oikawa, M. Shiraishi, Y. Suzuki, and K. Noguchi, Appl. Phys. Express {\bf 4}, 023003 (2011).
\bibitem{Ishikawa_PRB}
M. Ishikawa, T. Oka, Y. Fujita, H. Sugiyama, Y. Saito, and K. Hamaya, Phys. Rev. B {\bf 95}, 115302 (2017).
\bibitem{Jansen_PRAP}
A. Spiesser, H. Saito, Y. Fujita, S. Yamada, K. Hamaya, S. Yuasa, and R. Jansen, Phys. Rev. Appl. {\bf 8}, 064023 (2017).

\bibitem{Zhou_PRB}%
Y. Zhou, W. Han, L.-T. Chang, F. Xiu, M. Wang, M. Oehme, I. A. Fischer, J. Schulze, R. K. Kawakami, and K. L. Wang, Phys. Rev. B {\bf 84}, 125323 (2011).
\bibitem{Fujita1}
Y. Fujita, M. Yamada, S. Yamada, T. Kanashima, K. Sawano, and K. Hamaya, Phys. Rev. B {\bf 94}, 245302 (2016).
\bibitem{Fujita2}
Y. Fujita, M. Yamada, M. Tsukahara, T. Oka, S. Yamada, T. Kanashima, K. Sawano, and K. Hamaya, Phys. Rev. Appl. {\bf 8}, 014007 (2017).
\bibitem{Yamada_APEX}
M. Yamada, M. Tsukahara, Y. Fujita, T. Naito, S. Yamada, K. Sawano, and K. Hamaya, Appl. Phys. Express {\bf 10}, 093001 (2017).
\bibitem{Naito_APEX}
T. Naito, M. Yamada, M. Tsukahara, S. Yamada, K. Sawano, and K. Hamaya, Appl. Phys. Express {\bf 11}, 053006 (2018).

\bibitem{Peterson_PRB} 
T. A. Peterson, S. J. Patel, C. C. Geppert, K. D. Christie, A. Rath, D. Pennachio, M. E. Flatt\'e, P. M. Voyles, C. J. Palmstr\o m, and P. A. Crowell, Phys. Rev. B {\bf 94}, 235309 (2016).
\bibitem{Hamaya}
K. Hamaya, Y. Fujita, M. Yamada, M. Kawano, S. Yamada, and K. Sawano, J. Phys. D: Appl. Phys. {\bf 51}, 393001 (2018).

\bibitem{Zutic_RMP}%
I. \v{Z}uti\'c, J. Fabian, and S. D. Sarma, Rev. Mod. Phys. {\bf 76}, 323 (2004).
\bibitem{Dery_Nature}
H. Dery, P. Dalal, \L. Cywi\'nski, and L. J. Sham, Nature (London) {\bf 447}, 573 (2007).
\bibitem{Bratkovsky}
A. M. Bratkovsky, Rep. Prog. Phys. {\bf 71}, 026502 (2008).
\bibitem{Jansen}
R. Jansen, Nat. Mater. {\bf 11}, 400 (2012). 
\bibitem{Ramsteiner}
R. Farshchi and M. Ramsteiner, J. Appl. Phys. {\bf 113}, 191101 (2013).
\bibitem{Ciorga}
M. Ciorga, J. Phys. Condens. Matter. {\bf 28}, 453003 (2016).


\bibitem{Mattana_GaAs}
R. Mattana, J.-M. George, H. Jaffr\`es, F. N. van Dau, A. Fert, B. L\'epine, A. Guivar\'ch, and G. J\'ez\'equel, Phys. Rev. Lett. {\bf 90}, 166601 (2003).
\bibitem{Appelbaum_Si}
I. Appelbaum, B. Huang, and D. J. Monsma, Nature (London) {\bf 447}, 295 (2007). 
\bibitem{Hamaya_PRBR}%
K. Hamaya, M. Kitabatake, K. Shibata, M. Jung, M. Kawamura, S. Ishida, T. Taniyama, K. Hirakawa, Y. Arakawa, and T. Machida, Phys. Rev. B {\bf 77}, 081302(R) (2008).
\bibitem{Sasaki_APL}%
T. Sasaki, T. Oikawa, T. Suzuki, M. Shiraishi, Y. Suzuki, and K. Noguchi, Appl. Phys. Lett. {\bf 98}, 262503 (2011).

\bibitem{Ciorga_AAD}%
M. Ciorga, C. Wolf, A. Einwanger, M. Utz, D. Schuh, and D. Weiss, AIP Adv. {\bf 1}, 022113 (2011).
\bibitem{Bruski_APL}%
P. Bruski, Y. Manzke, R. Farshchi, O. Brandt, J. Herfort, and M. Ramsteiner,  Appl. Phys. Lett. {\bf 103}, 052406 (2013).
\bibitem{PLi_PRL}%
P. Li, J. Li, L. Qing, H. Dery, and I. Appelbaum, Phys. Rev. Lett. {\bf 111}, 257204 (2013).
\bibitem{Sasaki_APL2014}%
T. Sasaki, T. Suzuki, Y. Ando, H. Koike, T. Oikawa, Y. Suzuki, and M. Shiraishi, Appl. Phys. Lett. {\bf 104}, 052404 (2014).
\bibitem{Saito_JAP}
Y. Saito, T. Tanamoto, M. Ishikawa, H. Sugiyama, T. Inokuchi, K. Hamaya, and N. Tezuka, J. Appl. Phys. {\bf 115}, 17C514 (2014).
\bibitem{Kawano_PRmat}
M. Kawano, M. Ikawa, K. Santo, S. Sakai, H. Sato, S. Yamada, and K. Hamaya, Phys. Rev. Mater. {\bf 1}, 034604 (2017).
\bibitem{Oltscher_2DEG}
M. Oltscher, F. Eberle, T. Kuczmik, A. Bayer, D. Schuh, D. Bougeard, M. Ciorga, and D. Weiss, Nat. Commun. {\bf 8}, 18071 (2017).
\bibitem{Herfort_PRB}
Y. Manzke, J. Herfort, and M. Ramsteiner, Phys. Rev. B {\bf 96}, 245308 (2017).
\bibitem{Yamada_SST}
M. Yamada, T. Naito, M. Tsukahara, S. Yamada, K. Sawano, and K. Hamaya, Semicond. Sci. Technol. {\bf 33}, 114009 (2018).


\bibitem{TakahashiMaekawa}%
S. Takahashi and S. Maekawa, Phys. Rev. B {\bf 67}, 052409 (2003).
\bibitem{Jedema_PRB}%
F. J. Jedema, M. S. Nijboer, A. T. Filip, and B. J. van Wees, Phys. Rev. B {\bf 67}, 085319 (2003).
\bibitem{Kimura_Metal}
T. Kimura and Y. Otani, J. Phys. Condens. Matter. {\bf 19}, 165216 (2007).
\bibitem{KimuraHamaya}
T. Kimura, N. Hashimoto, S. Yamada, M. Miyao, K. Hamaya, NPG Asia Mater. {\bf 4}, e9 (2012).

\bibitem{Yu}
Z. G. Yu and M. E. Flatt\'e, Phys. Rev. B  {\bf 66}, 201202(R) (2002). 

\bibitem{Jamet_PRB}
C. Zucchetti, F. Bottegoni, C. Vergnaud, F. Ciccacci, G. Isella, L. Ghirardini, M. Celebrano, F. Rortais, A. Ferrari, A. Marty, M. Finazzi, and M. Jamet, Phys. Rev. B  {\bf 96}, 014403 (2017). 
\bibitem{Pezzoli_SR}
S. D. Cesari, R. Bergamaschini, E. Vitiello, A. Giorgioni, and F. Pezzoli, Sci. Rep. {\bf 8}, 11119 (2018). 
\bibitem{Watzinger_NC}
H. Watzinger, J. Kuku\v{c}ka, L. Vuku\v{s}i\'{c}, F. Gao, T. Wang, F. Sch\"{a}ffler, J.-J. Zhang, and G. Katsaros, Nat. Commun. {\bf 9}, 3902 (2018).


\bibitem{Sawano_TSF}
K. Sawano, Y. Hoshi, S. Kudo, K. Arimoto, J. Yamanaka, K. Nakagawa, K. Hamaya, M. Miyao, and Y. Shiraki, Thin Solid Films {\bf 613}, 24 (2016).
\bibitem{MYamada_APL}%
M. Yamada, K. Sawano, M. Uematsu, and K. M. Itoh, Appl. Phys. Lett. {\bf 107}, 132101 (2015).
\bibitem{Ando_APL}
Y. Ando, K. Hamaya, K. Kasahara, Y. Kishi, K. Ueda, K. Sawano, T. Sadoh, and M. Miyao, Appl. Phys. Lett. {\bf 94}, 182105 (2009).
\bibitem{Kasahara_JAP} 
K. Kasahara, Y. Baba, K. Yamane, Y. Ando, S. Yamada, Y. Hoshi, K. Sawano, M. Miyao, and K. Hamaya, J. Appl. Phys. {\bf 111}, 07C503 (2012).

\bibitem{Inomata}
K. Inomata, N. Ikeda, N. Tezuka, R. Goto, S. Sugimoto, M. Wojcik, and E. Jedryka,  Sci. Technol. Adv. Mater. {\bf 9}, 014101 (2008).
\bibitem{Hono}
R. Shan, H. Sukegawa, W. H. Wang, M. Kodzuka, T. Furubayashi, T. Ohkubo, S. Mitani, K. Inomata, and K. Hono, Phys. Rev. Lett. {\bf 102}, 246601 (2009).

\bibitem{Hamaya_PRL}
K. Hamaya, H. Itoh, O. Nakatsuka, K. Ueda, K. Yamamoto, M. Itakura, T. Taniyama, T. Ono, and M. Miyao, Phys. Rev. Lett. {\bf 102}, 137204 (2009). 
\bibitem{SYamada_APL}	 
S. Yamada, K. Tanikawa, S. Oki, M. Kawano, M. Miyao and K. Hamaya, Appl. Phys. Lett. {\bf 105}, 071601 (2014).

\bibitem{Vlado1} 
Z. Nedelkoski, B. Kuerbanjiang, S. E. Glover, A. Sanchez, D. Kepaptsoglou, A. Ghasemi, C. W. Burrows, S. Yamada, K. Hamaya, Q. Ramasse, P. Hasnip, T. Hase, G. Bell, A. Hirohata, and V. Lazarov, Sci. Rep. {\bf 6}, 37282 (2016). 
\bibitem{Vlado2}
B. Kuerbanjiang, Y. Fujita, M. Yamada, S. Yamada, A. M. Sanchez, P. J. Hasnip, A. Ghasemi, D. Kepaptsoglou, G. Bell, K. Sawano, K. Hamaya, and V. K. Lazarov, Phys. Rev. B {\bf 98}, 115304 (2018).

\bibitem{Supplemental} 
See Supplemental Material at [URL inserted by publisher] for a fabrication process of lateral spin-valve devices used in this study. 

\bibitem{Spiesser_APL}%
A. Spiesser, Y. Fujita, H. Saito, S. Yamada, K. Hamaya, S. Yuasa, and R. Jansen, Appl. Phys. Lett. {\bf 114}, 242401 (2019).

\bibitem{Fert_PRB}
A. Fert and H. Jaffr\`es, Phys. Rev. B {\bf 64}, 184420 (2001).
\bibitem{Fert1} 
A. Fert, J.-M. George, H. Jaffr\`{e}s, and R. Mattana, IEEE Trans. Electron Devices {\bf 54}, 921 (2007). 
\bibitem{Fert2} 
H. Jaffr\`{e}s, J.-M. George, and A. Fert, Phys. Rev. B {\bf 82}, 140408(R) (2010). 

\bibitem{Salis_PRB} 
G. Salis, A. Fuhrer, R. R. Schlittler, L. Gross, and S. F. Alvarado, Phys. Rev. B {\bf 81}, 205323 (2010).
\bibitem{Salis_PRBR2} 
G. Salis, S. F. Alvarado, and A. Fuhrer, Phys. Rev. B {\bf 84}, 041307(R) (2011).
\bibitem{Hu}
Q. Q. Hu, E. G. Garlid, P. A. Crowell, and C. J. Palmstr$\o$m, Phys. Rev. B {\bf 84}, 085306 (2011).

\bibitem{Jansen_PRAP2}
R. Jansen, A. Spiesser, H. Saito, Y. Fujita, S. Yamada, K. Hamaya, and S. Yuasa, Phys. Rev. Appl. {\bf 10}, 064050 (2018).

\bibitem{Chantis2}
A. N. Chantis and D. L. Smith, Phys. Rev. B {\bf 78}, 235317 (2008).
\bibitem{Crooker_PRB}
S. A. Crooker, E. S. Garlid, A. N. Chantis, D. L. Smith, K. S. M. Reddy, Q. O. Hu, T. Kondo, C. J. Palmstr{\o}m, and P. A. Crowell, Phys. Rev. B {\bf 80}, 041305(R) (2009).
\bibitem{Ando_PRB}
Y. Ando, K. Kasahara, S. Yamada, Y. Maeda, K. Masaki, Y. Hoshi, K. Sawano, M. Miyao, and K. Hamaya, Phys. Rev. B {\bf 85}, 035320 (2012). 
\bibitem{Shiogai_PRB}
J. Shiogai, M. Ciorga, M. Utz, D. Schuh, M. Kohda, D. Bougeard, T. Nojima, J. Nitta, and D. Weiss, Phys. Rev. B {\bf 89}, 081307(R) (2014).

\bibitem{Yamada_PRB}
M. Yamada, Y. Fujita, M. Tsukahara, S. Yamada, K. Sawano, and K. Hamaya, Phys. Rev. B {\bf 95}, 161304(R) (2017).

\bibitem{Jansen_PRL}
P. LeClair, J. T. Kohlhepp, C. H. van de Vin, H. Wieldraaijer, H. J. M. Swagten, W. J. M. de Jonge, A. H. Davis, J. M. MacLaren, J. S. Moodera, and R. Jansen, Phys. Rev. Lett. {\bf 88}, 107201 (2002).
\bibitem{Yuasa_NM}
S. Yuasa, T. Nagahama, A. Fukushima, Y. Suzuki, and K. Ando, Nat. Mat. {\bf 3}, 868 (2004).

\bibitem{Valenzuela_PRL}
S. O. Valenzuela, D. J. Monsma, C. M. Marcus, V. Narayanamurti, and M. Tinkham, Phys. Rev. Lett. {\bf 94}, 196601 (2005).
\bibitem{Bruski3}
P. Bruski, S. C. Erwin,  J. Herfort, A. Tahraoui, and M. Ramsteiner, Phys. Rev. B {\bf 90}, 245150 (2014).

\bibitem{Spiesser}
A. Spiesser, Y. Fujita, H. Saito, S. Yamada, K. Hamaya, W. Mizubayashi, K. Endo, S. Yuasa, and R. Jansen, Phys. Rev. Appl. {\bf 11}, 044020 (2019).

\bibitem{Crooker}
S. A. Crooker, M. Furis, X. Lou, C. Adelmann, D. L. Smith, C. J. Palmstr$\o$m, and P. A. Crowell, Science {\bf 309}, 2191 (2005).
\bibitem{Schultz}
B. D. Schultz, N. Marom, D. Naveh, X. Lou, C. Adelmann, J. Strand, P. A. Crowell, L. Kronik, and C. J. Palmstr$\o$m, Phys. Rev. B {\bf 80}, 201309(R) (2009).

\bibitem{Dash_SR}
M. V. Kamalakar, A. Dankert, P. J. Kelly, and S. P. Dash, Sci. Rep. {\bf 6}, 21168 (2016).
\bibitem{Gurram_NC}
M. Gurram, S. Omar, and B. J. van Wees, Nat. Commun. {\bf 8}, 248 (2017).

\bibitem{Tsymbal_PRL}
E.Y. Tsymbal, A. Sokolov, I. F. Sabirianov, and B. Doudin, Phys. Rev. Lett. {\bf 90}, 186602 (2003).

\bibitem{Tsukahara_APEX}
M. Tsukahara, M. Yamada, T. Naito, S. Yamada, K. Sawano, V. K. Lazarov, and K. Hamaya, Appl. Phys. Express {\bf 12}, 033002 (2019).

\end{thebibliography}
\end{document}